\documentclass[10pt,letterpaper]{article}
\usepackage[top=0.85in,left=2.75in,footskip=0.75in]{geometry}

\usepackage{changepage}

\usepackage[utf8x]{inputenc}

\usepackage{textcomp,marvosym}

\usepackage{amsmath,amssymb}

\usepackage{cite}

\usepackage{nameref,hyperref}

\usepackage[right]{lineno}

\usepackage{microtype}
\DisableLigatures[f]{encoding = *, family = * }

\raggedright
\usepackage[aboveskip=1pt,labelfont=bf,labelsep=period,justification=raggedright,singlelinecheck=off]{caption}

\makeatletter
\renewcommand{\@biblabel}[1]{\quad#1.}
\makeatother

\date{}

\usepackage{lastpage,fancyhdr,graphicx}
\usepackage{epstopdf}

\fancyheadoffset[L]{2.25in}
\fancyfootoffset[L]{2.25in}

\usepackage{newfloat}
\DeclareFloatingEnvironment[name={Supplementary Figure}]{suppfigure}
\DeclareFloatingEnvironment[name={Supplementary table}]{supptable}

\begin{document}
\vspace*{0.2in}

\begin{flushleft}
{\Large

\textbf\newline{Twitter Activity Timeline as a Signature of Urban Neighborhood}
}
\newline
\\
Philipp Kats\textsuperscript{1},
Cheng Qian\textsuperscript{1, 2},
Constantine Kontokosta\textsuperscript{1,2},
Stanislav Sobolevsky\textsuperscript{1,*},
\\
\bigskip
\textbf{1} Center for Urban Science and Progress, New York University, Brooklyn, New York, United States of America
\\
\textbf{2} Tandon School of Engineering, New York University, Brooklyn, New York, United States of America
\\
\bigskip

%
%





Correspondence should be addressed: sobolevsky@nyu.edu

\end{flushleft}
\section*{Abstract}

Modern cities are complex systems, evolving at a fast pace. Thus, many urban planning, political, and economic decisions require a deep and up-to-date understanding of the local context of urban neighborhoods. This study shows that the structure of openly available social media records, such as Twitter, offers a possibility for building a unique dynamic signature of urban neighborhood function, and, therefore, might be used as an efficient and simple decision support tool. Considering New York City as an example, we investigate how Twitter data can be used to decompose the urban landscape into self-defining zones, aligned with the functional properties of individual neighborhoods and their social and economic characteristics. We further explore the potential of these data for detecting events and evaluating their impact over time and space. This approach paves a way to a methodology for immediate quantification of the impact of urban development programs and the estimation of socioeconomic statistics at a finer spatial-temporal scale, thus allowing urban policy-makers to track neighborhood transformations and foresee undesirable changes in order to take early action before official statistics would be available.


\section*{Introduction}
Modern cities are complex systems with multiple layers of activities and interactions. Cities evolve constantly, changing in multiple ways and on multiple scales every day. Because of that, it is hard to describe every aspect of the system in a simple and compact description \cite{batty2008size, bettencourt2010urbscaling, bettencourt2013origins, arcaute2013citybound}. Yet, policy makers, urban planners, and other stakeholders need reliable, quantitative, and timely assessments of the characteristics for different metropolitan areas and neighborhoods \cite{kontokosta_quantified_2015, townsend2013}. Such a tool can provide a better understanding of urban patterns and trends, and the resulting information can supplement decision-making and help support successful approaches and practices across different cities \cite{powell2007food, bettencourt2007growth, albeverio2008, sobolevsky2015scaling}.

	The primary challenge here is the lack of data to support a consistent, quantifiable metric that provides a comprehensive understanding of multiple layers of urban operations. There are massive urban data collections gathered by various agencies and companies, with the potential to resolve this obstacle \cite{batty2012, bettencourt2014bigdata}. However, they are often expensive, rarely updated, and involve many methodological and privacy-related concerns. With proliferation of information and communication technologies, a new source of big data has become available and is starting to be exploited for urban research, including credit card transactions\cite{sobolevsky2014mining, sobolevsky2014money, sobolevsky2016prism, shen2014, scholnick2013}, GPS traces\cite{santi2014, kung_exploring_2014}, transport card usage \cite{bagchi2005, lathia2012, chan1999frauddetection, rysman2007}, garbage transportation\cite{offenhuber_putting_2012}, 311 complaints \cite{wang_structure_2016} and mobile call records \cite{reades_cellular_2007, reades_eigenplaces:_2009, jacobs-crisioni_linking_2012, pei_new_2014, girardin2008digital, ratti_mobile_2006, gonzalez2008uih, quercia2010rse, ratti2010redrawing, sobolevsky2013delineating, amini2014impact,grauwin_towards_2015}.
    
	There is a significant limitation, however, in adopting such analytic methods in practice, given the extreme heterogeneity of data, as different sources are available for different areas and periods of time. Data collection methods also vary drastically. This, combined with logistical issues in accessing these data and privacy considerations, can make the widespread use of such data infeasible \cite{lane2014, christin2011,belanger2011, krontiris2010}. 
    
	Another data source gained a lot of attention recently and has become a popular subject of study: social media\cite{szell2013, jurdak_understanding_2015,cranshaw_livehoods_2012, frank2013happiness, hawelka2014, paldino2015, gordon_net_2011, lenormand_bcards_2014}. Combining large quantities of fine-grained information that is accessible in real time and with the global coverage, social media data, for example, Twitter, offer an important new source of data for urban studies. Obviously, Twitter has its own limitations when used for this purpose: it is still relatively sparse (although usage grows at a rapid pace and this limitation might soon become obsolete) and, more importantly, does not provide a complete representation of the urban population, as users tend to be younger and more educated. There might also be spatial biases involved as well, as people may act differently on social media according to their location \cite{jurdak_understanding_2015}. Last, the role and behavioral patterns of particular channels may vary across the globe; communities might use the same social media platform for different purposes based on cultural and political differences. Still, the opportunities of using social media for research purposes are unprecedented, especially those focused on learning patterns of social media usage over time and space, rather than trying to directly extrapolate it to represent the entire urban population.
    
In the present paper, we analyze the timelines of Twitter activity in New York City to study urban landscape decomposition, detect anomalous events, and model socioeconomic properties of identified neighborhoods. In section 1, we describe the data used and the necessary data manipulations and cleaning procedures. Section 2 covers the definition of the typical weekly twitter activity timeline signatures (TWS). In section 3, we use a set of constructed TWS for zip codes as a baseline for event detection and event impact estimation. In section 4, we apply clustering of TWS and investigate the difference in functional features of the resulting clusters. Finally, in sections 5, 6 and 7, we use machine learning algorithms to model those features and discuss the implications of our findings.

\section{Data Processing}

\subsection{Twitter}

\begin{figure}[ht!]
\centering
\includegraphics[width=130mm]{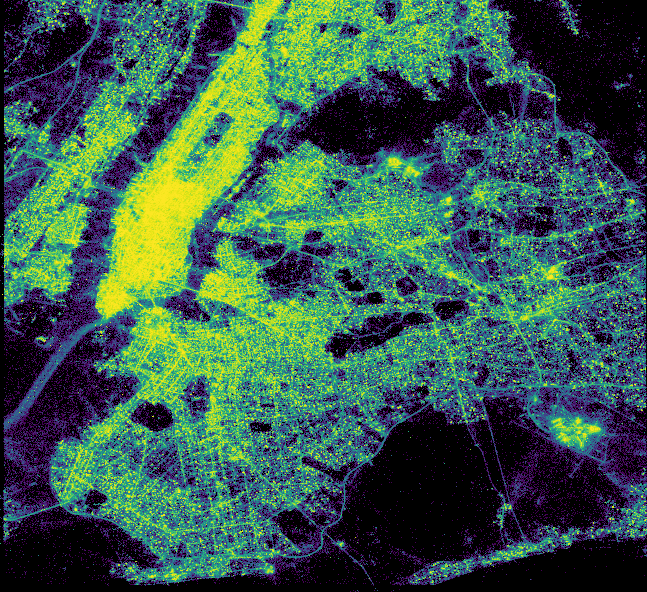}
\caption{Spatial distribution of tweets}
\label{fig1}
\end{figure}

Twitter is a popular micro-blogging service that allows people to post short messages and follow other people across the world. In the first quarter of 2016, the number of monthly active users worldwide exceeded 310 million. Due to its (recently removed) text size limitation, users tend to generate posts at a fast pace, through multiple native and third-party applications. Due to the platform's popularity, any approach based on its data is \textit{a priori} applicable to  most urban areas across the globe. With the collection of historical records, and detailed information on the particular time, user, application, post geographical coordinates and the body of message, Twitter has become a rich source of  information on the characteristics of urban landscape. 
A feed of tweets with geolocation from four boroughs (excluding Staten Island) in New York City was collected from January 2014 to June 2016, using Twitter's official API. These data were then aggregated to the 262 NYC Postal (Zip) Codes; tweets considered as automated (more on data processing in the Appendix I) were removed. Our final database contains over 10 million tweets from about 1,300,000 unique users. Those tweets were generated using 603 different applications, though 92\% were generated by Twitter native mobile applications, Instagram, and Foursquare, as presented in table \ref{table1}. The resulting spatial distribution of tweets is presented on fig.  \ref{fig1}.

\begin{table}[ht!]
\centering
\caption{Services distribution, top 5}
\label{table1}
\begin{tabular}{llll}
\textbf{Application Name}            & \textbf{users}  & \textbf{tweets}   &\textbf{ tweets,\%}\\
Twitter for iphone          & 449723 & 20081109 & 56.8425 \\
Instagram                   & 395185 & 5575630  & 15.7826 \\
Twitter for Android         & 139554 & 4823926  & 13.6548 \\
Foursquare                  & 123050 & 2506839  & 7.0960 \\
Twitter for ipad            & 15875  & 434874   & 1.2310 \\
\end{tabular}
\end{table}

\subsection{Neighborhood characteristics}
Additional data sources are used to define the functional properties of the urban areas, including land use data and social-economic characteristics from the US Census 2014 American Community Survey.

\subsubsection{Land Use data}
The New York City Primary Land Use Tax Lot Output dataset, more commonly known as PLUTO \cite{nyc_department_of_city_planning_pluto}, maintains land use, zoning, and property characteristics for more than 1,100,000 tax (property) lots in the City. Each lot has a corresponding postal code, land use code and area of the lot, among many other variables. We aggregate these data for each zipcode to classify the land use context of each area, such as residential areas with low-density buildings, or commercial districts with multi-story office buildings. New construction activity can quickly change the demographics and mobility patterns of an area, thus affecting the patterns of Twitter usage. 

\subsubsection{2014 US Census data}

The 2014 American Community Survey \cite{bureau_american_2014}, the latest at the moment, contains hundreds of demographic and economic characteristics for each zipcode. We select relevant features to describe individual neighborhoods, including population diversity, education, wealth, unemployment rates, etc. In particular, we use the number of population in the following categories: "Non-Hispanic White", "African American", "Asian", "High school degree", "College degree", "Graduate degree", "Uninsured ratio", "Unemployment ratio", "Poverty ratio", and mean for "Income (all)", "Income of No Family", "Income of Families" and "Income of Households". More information on data sources can be found in SI \nameref{target}.

\subsection{Spatial units}

Another important consideration for our research is selecting the appropriate spatial scale. For the purpose of the present study, we made a decision to aggregate the data at the zip code level, which can account for neighborhood or sub-neighborhood boundaries. It is worth noting that the Twitter dataset contains accurate (up to 10 meters) GPS coordinates of each tweet, and it can be aggregated with a great flexibility to almost any spatial granularity. While smaller units of aggregation require larger quantities of data to keep estimations statistically significant, this flexibility is, in our opinion, one of the large advantages of using Twitter data to approximate the local urban context over other sources, enabling scalable solutions prepared to work at a finer spatial scale once the data density becomes sufficient.

\section{Typical temporal variation of Twitter activity for urban locations}

It is expected that different urban areas will demonstrate different patterns of Twitter activity, based on their social and economic context. In order to understand regular behavioral patterns for each neighborhood, we define the typical weekly activity timeline signature (TWS) for each zipcode, each based on twitter transactions, following the procedure presented in \cite{reades_cellular_2007a}. 

	TWS is presented by an array of N quantities, where N denotes the number of time bins considered within an average week. While the length of such period can vary, in this research we consider two type of bins - 15 min bins for the average weekly activity analysis and 6 hour bins for the event detection timeline, as we are working with the specific week in this case. For each bin \textbf{\textit{i}}, we calculate the average number of tweets \boldmath{$\mu[i]$} falling into the respective bin across the entire dataset  (for example, all collected Tweets, published within time range from 15:00 to 15:15 on any Monday), and then normalize it by the total number of tweets across all the bins (see formula \ref{formula1}):

\begin{equation}
\centering
T[i] = \frac{\mu[i]}{\sum_{i=1}^N \mu[i]}
\label{formula1}
\end{equation}

	A corresponding TWS is generated for the City in total and each zip code individually. Finally, we generate TWS for each of the top four platform applications. In order to stay free of the outliers largely caused by the data sparsity, the 15Min TWS with at least one bin, containing more than 10\% of the total amount of tweets are dropped. 
	TWS represents a typical pattern of Twitter-related activity in a particular zip code. It serves as a unique local signature resulting from the general city-wide trends as well as particular location-specific patterns.

\begin{figure}[ht!]
\centering
\includegraphics[width=130mm]{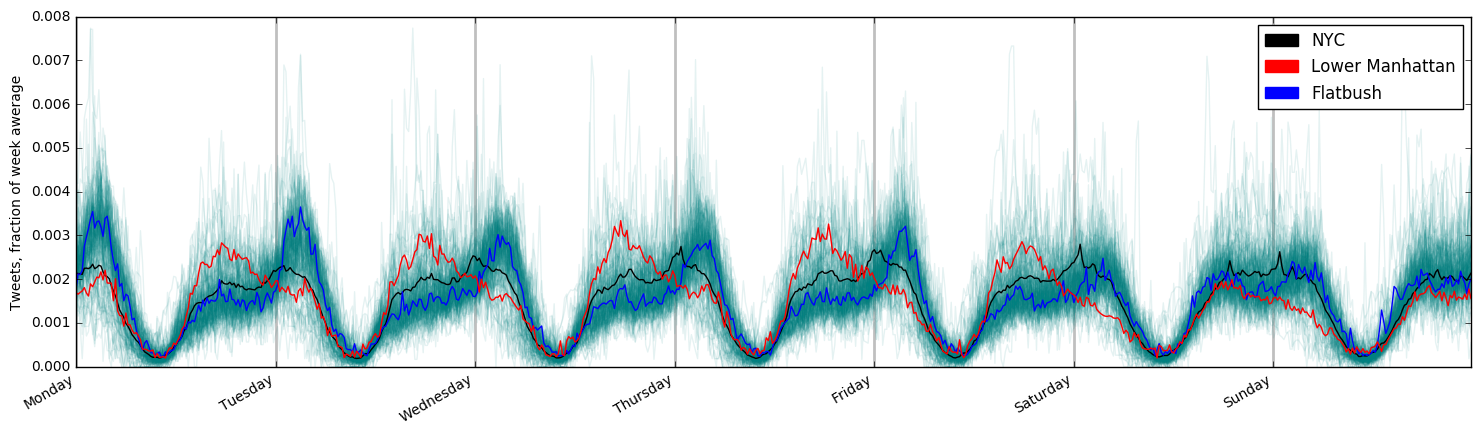}
\caption{Typical week, New York City, threshold}
\label{fig3}
\end{figure}

\begin{figure}[ht!]
\centering
\includegraphics[width=130mm]{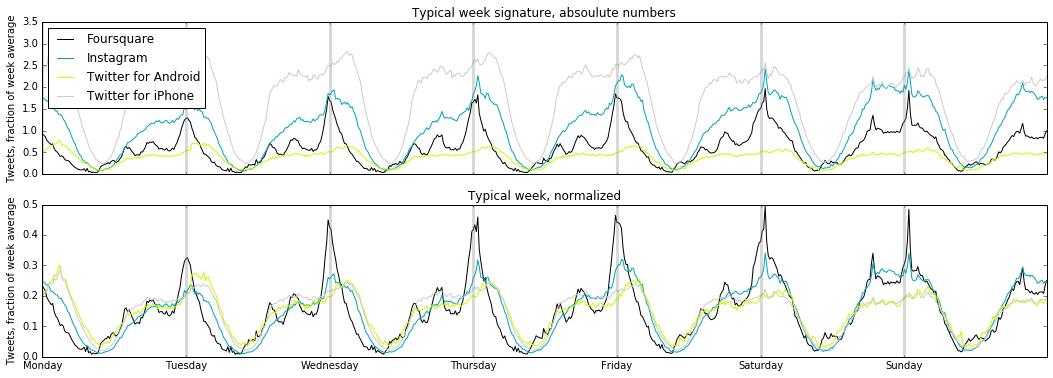}
\caption{Typical week, New York City, 4 Top Applications}
\label{fig4}
\end{figure}

Typical signatures for the city in the aggregate and two specific neighborhoods — Lower Manhattan, the heart of New York's downtown commercial district, and Flatbush, a residential area in Brooklyn - are presented on figure \ref{fig3}. TWS for all zip codes are also presented as background to provide some general context on how twitter use varies spatially and temporally. 

As can be seen, local patterns are very distinctive between different neighborhoods in the City. The midnight peak is only observed for the overall NY and Flatbush, while Lower Manhattan, representing a business district, has significant activity during a typical workday. The pattern changes slightly throughout the week, with afternoon/evening peaks later on Fridays and earlier on Sundays and Mondays. 

Given these observed patterns, it would be reasonable to consider the hypothesis that the TWS could serve as a unique indicator of urban function and structure, and that neighborhoods with similar TWS share other similar characteristics. Building from this proposition, the TWS could be applied to detect and indicate anomalous or atypical events, a possibility that will be introduced in the next section.

\section{Event detection}

In January 2015 (23-31), a powerful blizzard affected the Eastern United States. On Monday, January 25, the storm approached New York City and the following day brought many road closures and other transit disruptions. As the geocoded Twitter stream represents human behavior and mobility, one would expect typical patterns to be affected by such a major weather event. Using this opportunity, we investigate the impact of significant exogenous events on the Twitter TWS. In this case, as data for a particular week is sparse, we decide to switch to a 6-hour range to calculate zip code TWS, and compare differences between activity in an average week against the particular week of the storm.

\begin{figure}[ht!]
\centering
\includegraphics[width=90mm]{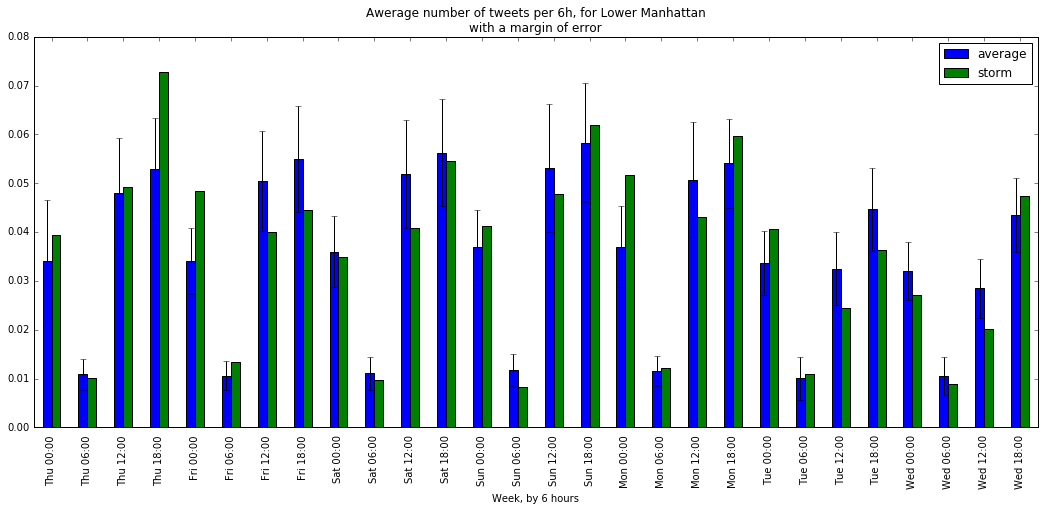}
\caption{Storm week versus typical week comparison for Lower Manhattan}
\label{fig7}
\end{figure}

\begin{figure}[ht!]
\centering
\includegraphics[width=90mm]{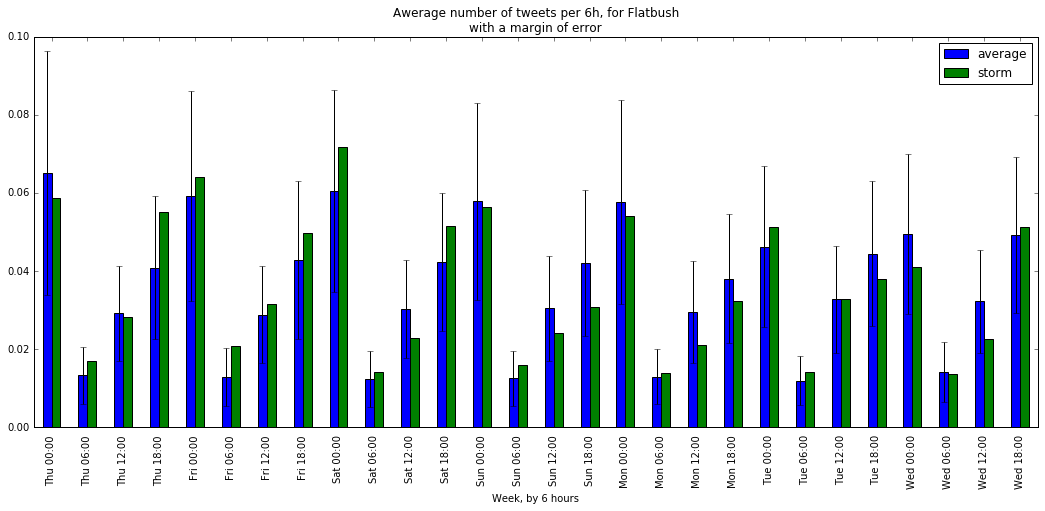}
\caption{Storm week versus typical week comparison for Flatbush}
\label{fig8}
\end{figure}

Indeed, as figures \ref{fig7} and \ref{fig8} show, there is a significant shift in patterns for the two days of the storm event, January 25 and 26. The extent and direction of pattern changes, however, differs depending on the specific area: for Lower Manhattan, Twitter activity was significantly higher on Monday evening, and significantly lower during Tuesday work hours. On the contrary, in Flatbush, neighborhood Twitter activity was higher during work hours on Monday and Tuesday evening, and saw a spike on Tuesday morning. This fact might be explained by the attention to the event and the severe weather and bad road conditions: initially, higher activity in business areas before people returned home from work and started taking the storm seriously, and then later, lower activity in business areas and increased activity in residential areas as many did not make it to work next day.

\section{Clustering urban locations through temporal activity signatures}

The difference in temporal signatures of locations might be related to differences in their urban function. Indeed, we can assume that activity patterns can be dependent on the average work hours, commute times and modes, wealth, and education of the neighborhood. If true, decomposition of the TWS might be used to reveal major areas of similar or distinctive functional profiles across the city, and serve as a first proof of concept for further modeling of socio-economic characteristics using Twitter patterns.

To decompose neighborhoods by their typical weekly signature, we perform clustering using the k-means algorithm, chosen as a well-known, robust and fast technique. In order to choose the most appropriate number of clusters, we use the Silhouette score and Elbow approaches. Please see further details on the clustering methodology in SI \nameref{clustering}. 

As we discussed earlier, TWS might represent a mixture of multiple trends and area characteristics at different scales. Because of that, different clusters may represent different scope of trends: for example, we achieve the best Silhouette score with $k=2$. This number of clusters leads to a partition, easily interpretable both from the time series plot and from the map, that represents areas with more and less active night life, e.g. downtown areas, cultural districts and airports versus residential areas. With the increase of k, the clusters becomes more detailed, adding more aspects to the partitioning: with $k=3$, we can see how areas are split into the dense downtown/commercial districts, residential zones, and areas of the mixed uses, including Williamsburg, Brooklyn and areas surrounding downtown Brooklyn. Partitioning for $k=5,7$ provide further refinement to differences in residential areas, introducing new clusters. Here, one might identify the split of low-story residential areas into two large sub-clusters. Another split defines dense downtown areas and less commercial adjacent neighborhoods.  The Elbow approach, on the other hand, suggests to use $k=7$ as the most appropriate number. As we are interested in having additional detail and more fine-grained clustering of urban locations, partitioning of 7 clusters seems to be preferable.

\begin{figure}[ht!]
\centering
  \includegraphics[width=130mm]{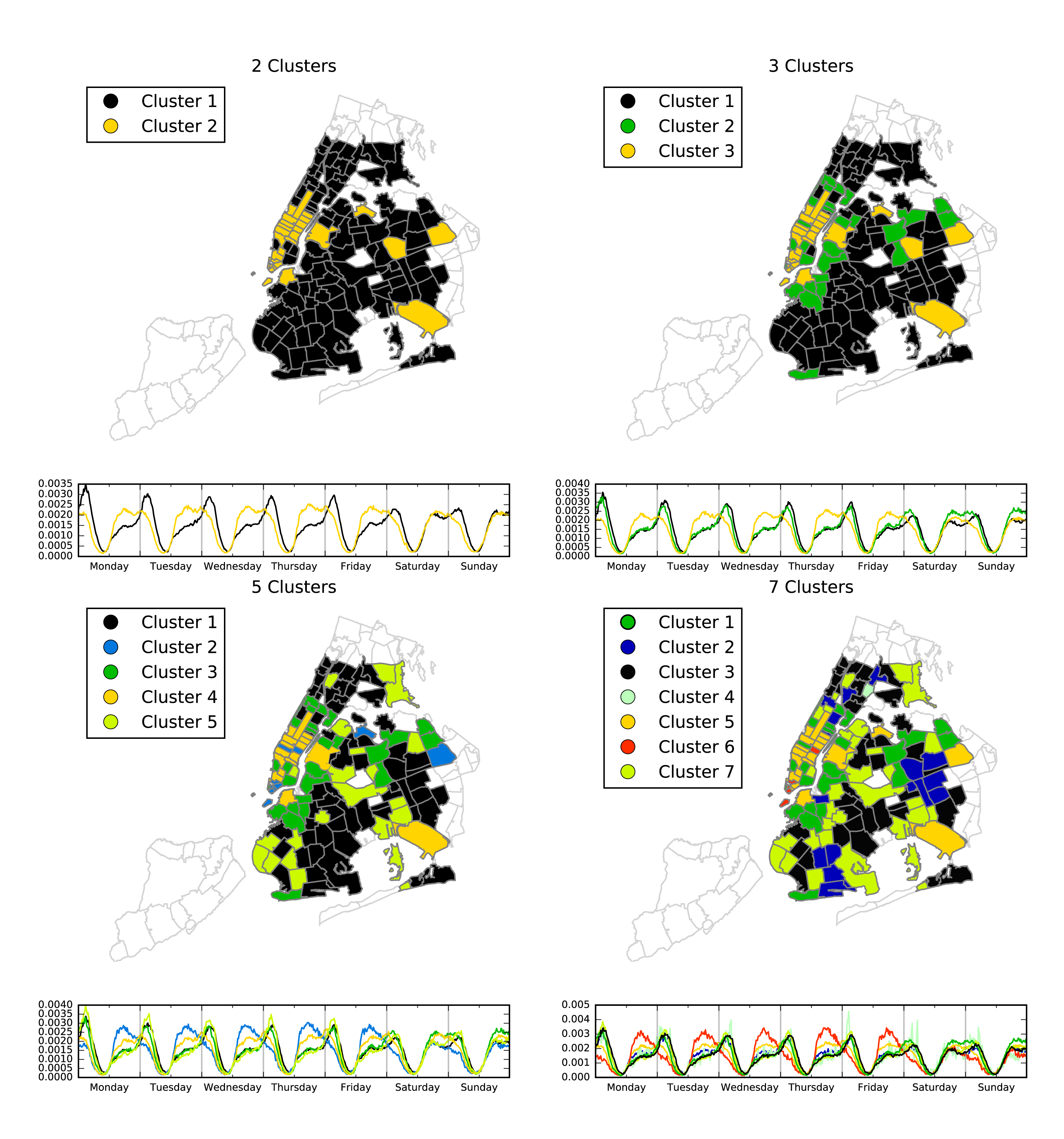}
\caption{Partition of TWS for 2, 3, 5 and 7 clusters}
\label{clusters_plot}
\end{figure}

\subsection*{Neighborhood properties decomposition}

\begin{figure}[ht!]
\centering
\includegraphics[width=130mm]{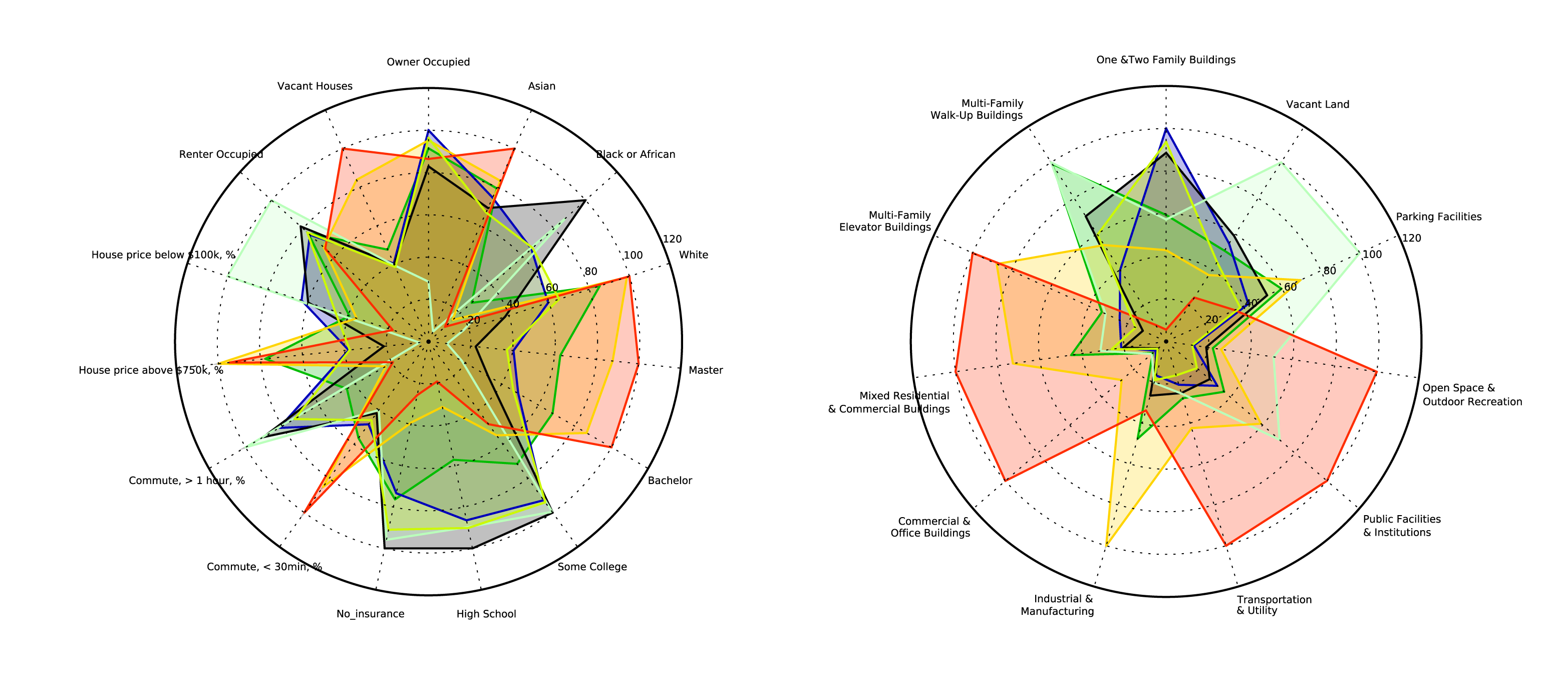}
\caption{Land use, Socioeconomic characteristics for 7 clusters, Percentage of maximal value}
\label{fig12}
\end{figure}

Next, we quantitatively interpret the output of the cluster partitioning. Recall that those results are obtained with Twitter stream TWS using k-means clustering, with no spatial information considered. We then compare aggregated socio-economic characteristics of each cluster. The result for Land use and Demographics are plotted on the fig. \ref{fig12}. Specifically, we compare clusters by the percentage of different land use types, as well as the average values for a set of relevant social-economic features, including percentage of inhabitants by race, education, health insurance coverage,average income, house prices, and commute times. While some particular properties differ only slightly, general split between residential, mixed and commercial areas, as well as  zip codes with large recreational zones (Central Park, for example), is clear. Only a few socioeconomic parameters of the areas correlate with TWS, but land use fits time series well. With that knowledge, it seems reasonable to model area's functional zoning with TWS.

Plots represent significant differences between the clusters, revealing strong connections between Twitter TWS and an area's land use and social properties. This finding serves to support our hypothesis that Twitter patterns can be used to characterize local context. While the clustering reveals a general relationships, which are easy to interpret, but hard ones to use in on practice. At the same time, having those results we can expect more detailed relations to exist. In particular, it rises another question: can we apply TWS approach to model socioeconomic properties on the local scale? 

\section{Modeling the functional properties}

As we discussed earlier, the Twitter TWS allows us to aggregate city zip codes into meaningful clusters, quite distinctive in their functional patterns. Given this, we can ask if Twitter can then be used to model and estimate those functional characteristics. If yes, such a model could be useful, for example, in order to improve our understanding of changes to local land use, as the area's use may be transformed at a rapid pace. Creating a reliable model based on publicly-available data being updated in near-real-time has a substantial potential for application in various domains.

We apply a regression model using TWS for each zip code as the feature space (see Appendix for details). Using this approach, we are trying to model six target variables, including percentage of One- and Two-Family Buildings, Commercial and Office Buildings, Open Space and Outdoor Recreation Area, Multifamily Elevator Buildings, percentage of Owned Houses, and percentage of inhabitants having commutes of less than 30min. All characteristics here are extracted from the PLUTO dataset and American Community Survey.

Our objective here is to use partial information about the target variables defined in a certain part of the city to train the model so the one can explain the target variables over the rest of the city. In order to avoid over-fitting and find the optimal hyper-parameters for each model, the data are split into 3 parts — training, test and validation parts, represented by 60\%, 20\% and 20\% of the records.

Next, the model is trained using the training set including the corresponding values for the target variable. Using grid search, we apply different combinations of appropriate hyper-parameters - kernel types, coefficients, number of trees, etc., evaluating performance over the validation set. Finally, the model with the best set of parameters is applied to the test set for the final evaluation.

Four models are applied, including Random Forest Regression, Neural Networks with regularization (NN), Extra Trees Regression (ETR) and XGBoost algorithm. \cite{breiman_random_2001, girosi_regularization_1995, geurts_extremely_2006, chen_xgboost:_2016}. 
We use R\textsuperscript{2} as our objective criteria.  In general, Extra Trees on average performed better than other models, avoiding over-fitting and resulting in better R\textsuperscript{2} both on training/validation and test sets (out of sample performance). The resulting out of sample R\textsuperscript{2} values for the best models selected are summarized in Table \ref{tws_model}. Technical details on the models can be found in SI \nameref{modeling}.

\begin{table}[ht!]
\centering
\caption{Modeling performance for the top 10 features}
\label{tws_model}
\begin{tabular}{lllll}
                                & \multicolumn{2}{l}{overall TWS}                              & \multicolumn{2}{l}{app-specific TWS}     \\
\textit{\textbf{feature}}       & \textit{R\textsuperscript{2}} & \textit{\textbf{best model}} & \textit{R\textsuperscript{2}} & \textit{\textbf{best model}}\\
Commercial \& Office Buildings, \%  & 0.6902                    & nn                            & 0.7312             & nn                  \\
Commute, less than an hour     & 0.6035                    & xgb\_gbdart                   & 0.6345             & xgb\_gbdart         \\
population with Master's degree                          & 0.574289                    & et                            & .7264              & xgb\_gbdart         \\
House price above \$750k             & 0.5323                    & rt                            & 0.5477             & et                  \\
African American population                & 0.5322                    & nn                            & 0.3185             & rf                  \\
population with Bachelor's degree                        & 0.5166                    & xgb\_gbdart                   & 0.5366             & et                  \\
population with High School                     & 0.4930                    & et                            & 0.4079             & et                  \\
population with College education                    & 0.3924                    & et                            & 0.4363             & rf                  \\
Owner Occupied houses                  & 0.3533                    & xgb\_gbdart                   & 0.2234             & et                  \\
One\&Two Family Buildings      & 0.3238                    & et                            & 0.3564             & xgb\_gbdart         \\
Commute below 30 Minutes       & 0.2548                    & et                            & 0.37               & rt                    
\end{tabular}
\end{table}

We see that certain target variables appear to be modeled better than the others. In particular, we are especially effective in predicting commute time and percentage of Commercial Buildings. It is most likely that, as we have seen with the clustering before, those parameters have the highest impact on the typical weekly activity. It would be logical to consider other transportation and behavioral characteristics in the future, e.g. number of employees for specific industries, number of restaurants and coffee shops, etc.

There are a few reasons to consider these modeling results as important. First, they indicate a strong relationship between Twitter stream patterns and some of the local urban properties. Second, they enable other potential predictions of local land use features, revealing corresponding changes and trends, which might be even more important given the long update cycles for land use records. Third, such models can be potentially applied to any arbitrary spatial scale, leading to a more fine-grained analysis, for example, at the block level. 

\section{Detailed Service-based timeline}

Each tweet, obtained through the Twitter API, contains metadata, including the name of the particular service that was used to generate this tweet. Services represent different functions, and each of them has a unique timeline: Foursquare, for example, has clear peaks during brunch, lunch and dinner times. It is reasonable to assume that those timelines introduce additional information for the urban areas. In this case, we treat TWS for the four most popular services as a single set of features, that contains 2,688 features. The model, trained on the new dataset performs substantially better than the first one (see fig. \ref{tws_model}). 

This detail level, however, comes at a price: while resulting in better modeling results, the approach is more computationally demanding and requires more data to be fed for each area. Because of that, it is generally more limited to the larger and more populated areas. For our dataset, the number of areas decreased significantly: while for ordinary TWS we use 135 zip codes, the latter approach works only for 96 of them, given 4 applications. On top of that, large set of features increases the chances of overfitting, which we have to avoid using the test set of data. With that, there is a trade-off between the performance and the data required, as one can change different number of services to include in the model. 

\section*{Conclusion}

The understanding of constantly-evolving local urban context is crucial for urban management, planning and decision-making. In this paper, we have considered possible approaches to quantify local signatures of urban function through a stream of records from Twitter within New York City. Similar data feeds can be easily collected for most of the major cities across the world, and scaled to different levels of spatial and temporal resolution. Particular limits of scale, however, are yet to be defined. As this study shows, there is a strong and consistent relationship between certain characteristics of urban areas and the corresponding observed activity of social media users. 

For the analyzed dataset, we demonstrate how similarities in TWS can produce meaningful urban clusters characterized by distinctive socioeconomic patterns. In particular, we were able to define 7 clusters of zipcodes, representing areas of active night life (including airports and commercial downtown cores), low-rise and hi-rise residential areas, mixed-use areas, etc. The very same TWS has been used to detect major events and estimate neighborhood exposure. Indeed, we were able both to detect one of the recent snow storms and to estimate the neighborhoods most affected by the hazard. The full set of time series for different locations allowed us to apply regression models to analyze the functional characteristics of urban neighborhoods, such as commute times, number of owned houses and commercial land uses with sufficient accuracy. In particular, out-of-sample accuracy of the random forest model explaining the fraction of people with graduate degrees, college diplomas, average commute time, and the proportion of higher-priced houses can be modeled with R\textsuperscript{2} values of .55, .54, .47 and .39 correspondingly. For the same target quantities the Extra Trees Regression model performs with the corresponding out-of-sample R\textsuperscript{2} scores of 0.44, 54, .47 and .65. 

As the Twitter stream contains a source for the tweet, e.g. a label of the specific application the particular message has been generated from, this makes it possible to construct a more detailed pattern. Using TWS for the top popular applications - Instagram, Foursquare, and Twitter application for iPhone and Android - we create a more detailed feature space of 2,688 features in total. This new feature set improves the results: the corresponding model performed achieves a 0.62 R\textsuperscript{2} for the proportion of commercial property in a given zip code. However, this approach requires sufficient data for each feature to be adequately defined, and thus has more strict limitations on the scale of the area considered.

Twitter is already being leveraged in many ways for urban analysis, including fraud detection, protests, human mobility inference, and others. It is likely that this type of source will start being used to estimate neighborhood characteristics among urban areas, supporting informed policy making and planning decisions. 

While proper modeling requires large quantities of data, social media is increasingly dense and relatively accessible for any large city in the world. With sufficient data, one could imagine developing a set of efficient instrumental variables based on different kinds of social media. This set, then, leveraged by the appropriate socio-economic models, can be used to substitute official socioeconomic statistics and the functional properties of the areas where official data are non-existent or inconsistent. By developing new sources for extracting localized information an greater temporal resolution, these models can help to quantify and evaluate impacts from urban interventions, emergencies, or disasters, causing a significant shift in how urban analysis can be conducted. 
\nolinenumbers

\bibliography{biblio.bib}

\section*{Supplementary materials}

\subsection*{Data collection and preparation} \label{data_processing}

All tweets were collected through official Twitter API data provider. In total, more than 10 million of tweets within the boundaries of New York City were stored since 2015 as well as some earlier periods. Tweets were then geocoded to New York City zipcode boundaries, so that each tweet received a label of zipcode it belongs to.

\subsection*{Spam detection}\label{spam}
	Some applications generate tweets automatically and do not represent any human activity due to their automatic nature. Many of them generate significant quantities of messages during specific short time periods. Because of that, they can potentially impact time series, and should be removed. It was noted that many of those applications have a large ratio of tweets per user, while being represented by very small number of user ids. Empirically we know that most of them have less than a dozens of users, each of whom generate thousands of messages a day. Therefore, we drop all the tweets, generated by any applications, for which two conditions are met: First, more than 5\% of total tweets belongs to one single user, and second, total number of tweets for one user is larger than 1000. Using this technique, 12 particular application are dropped, including NYC\_511 road traffic bot, NYC job offer bot, and a few others, responsible for about 0.5\% of the total database. All removed applications are present in Table \ref{spamtable}.

\begin{supptable}[ht!]
\centering
\caption{Applications, which activity considered as automated.}
\label{spamtable}
\begin{tabular}{llllll}
application          & users & tweets & tweets,\%   &  &  \\
511NY-Tweets         & 12        & 26759 & 0.000757 &  &  \\
pbump.net            & 2         & 19763 & 0.000559 &  &  \\
twanoniem            & 1         & 18598 & 0.000526 &  &  \\
twitrriffic          & 1         & 16591 & 0.000470 &  &  \\
Goldstar             & 14        & 15662 & 0.000443 &  &  \\
COS App              & 1         & 13614 & 0.000385 &  &  \\
kickalert            & 1         & 10249 & 0.000290 &  &  \\
Weavrs               & 7         & 8754  & 0.000248 &  &  \\
dine here            & 1         & 8629  & 0.000244 &  &  \\
SimplyCast           & 5         & 7277  & 0.000206 &  &  \\
Tide Bot             & 2         & 6635  & 0.000188 &  &  \\
screamradius         & 1         & 5457  & 0.000154 &  &  \\
UNjobs               & 16        & 4006  & 0.000113 &  &  \\
TTN NYC traffic      & 1         & 3607  & 0.000102 &  &  \\
Ivory Standard       & 1         & 3471  & 0.000098 &  &  \\
Vinny Scans          & 1         & 2995  & 0.000085 &  &  \\
eLobbyist            & 1         & 2816  & 0.000080 &  &  \\
Yakaz                & 3         & 2789  & 0.000079 &  &  \\
Skiplagged Deals     & 1         & 2358  & 0.000067 &  &  \\
TownTweet            & 5         & 1978  & 0.000056 &  &  \\
Dexigner iPhone App  & 2         & 1626  & 0.000046 &  &  \\
Words \& Warps       & 2         & 1457  & 0.000041 &  &  \\
511NY                & 6         & 1332  & 0.000038 &  &  \\
TwittlyDumb          & 1         & 1256  & 0.000036 &  &  \\
Authentic Jobs       & 1         & 1108  & 0.000031 &  &  \\
Tweetings for iPhone & 12        & 1093  & 0.000031 &  & 
\end{tabular}
\end{supptable}

\subsection*{Neighborhood characteristics}\label{target}

Two major sources were used in order to describe neighborhood characteristics - NYC Tax Lot database  (PLUTO) and US Census 2014. Fig. \ref{table_land_use} represent overall distribution of land use per lot on average in NYC.
\begin{supptable}[ht!]
\centering
\caption{Land use, NYC}
\label{table_land_use}
\begin{tabular}{ll}
\textbf{Land use permission type}            & \textbf{\%, NYC}\\

One \& Two Family Buildings                & 27.3\% \\
Multi - Family Walk- Up Buildings          & 7.2\%  \\
Multi - Family Elevator Buildings          & 5.2\%  \\
Mixed Residential and Commercial Buildings & 3.2\%  \\
Commercial and Office Buildings            & 3.9\%  \\
Industrial and Manufacturing               & 3.5\%  \\
Transportation and Utility                 & 7.5\%  \\
Public Facilities and Institutions         & 6.9\%  \\
Open Space and Outdoor Recreation          & 27.2\% \\
Parking Facilities                         & 1.4\%  \\
Vacant Land                                & 6.5\% 
\end{tabular}
\end{supptable}

The 2014 census data contains hundreds of features. Based on our common understanding, we selected 15: Percentage of Owner Occupied Houses, Vacant Houses, Renter Occupied, Houses with price below	100.000	 or above 750.000, residents with daily commute above an hour or less than 30 min, residents with no health	insurance, people with High School, College, Bachelor or Master's degree as a higher education available, percentage of population of different races. All characteristics are presented on table \ref{census_nyc}.

\begin{supptable}[tbp]
\centering
\caption{Chosen characteristics from Census, and their percentage for the city as a whole}
\label{census_nyc}
\begin{tabular}{ll}
Characteristic                   & NYC       \\
Owner Occupied                   & 26.382030 \\
Vacant Houses                    & 9.447669  \\
Renter Occupied                  & 64.170302 \\
house price \textless100         & 4.836802  \\
house price\textgreater750.000   & 25.720689 \\
Commute, hour or more            & 22.035100 \\
Commute, less than 30 min        & 30.923152 \\
No Health Insurance Coverage     & 7.254104  \\
High School                      & 23.724852 \\
Some College                     & 19.809675 \\
Bachelor                         & 21.213632 \\
Master                           & 10.129234 \\
White                            & 43.496439 \\
Black or African                 & 24.348927 \\
Asian                            & 14.125574
\end{tabular}
\end{supptable}

\subsection*{Clustering}\label{clustering}

In order to define number of clusters, two particular techniques were used: Silhouette score and an Elbow method (fig. \ref{silhouette}, \ref{elbow}). Silhouette score measures how on average each data point is closer to its cluster's center than to any other cluster. Elbow method represents cumulative square error (CSE) for all points within each cluster. Having the plot of CSE for different numbers $k$ of clusters, we can select the optimal $k$ at to the point, where "growth" of the CSE slows down.

\begin{suppfigure}[ht!]
\centering
\includegraphics[width=60mm]{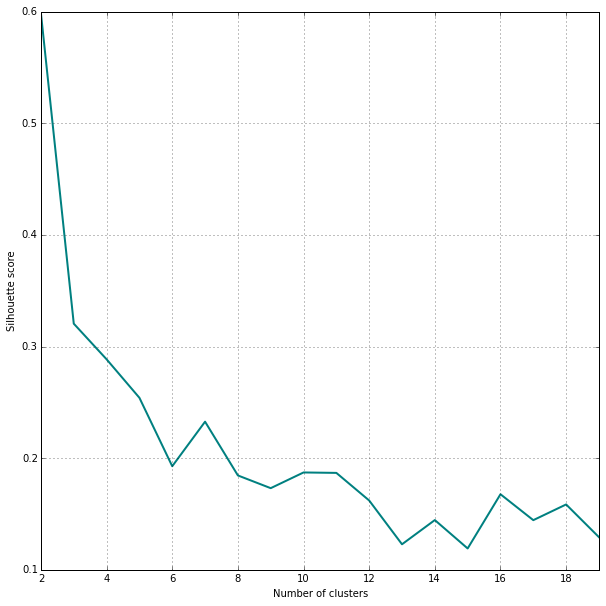}
\caption{TWS K-mean clustering Silhouette scores for different {k}, overall }
\label{silhouette}
\end{suppfigure}

\begin{suppfigure}[ht!]
\centering
\includegraphics[width=60mm]{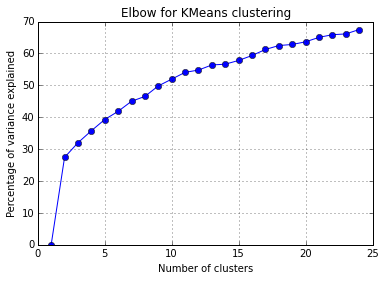}
\caption{Elbow method for k-means }
\label{elbow}
\end{suppfigure}

\subsection*{Regression Modeling}\label{modeling}

In order to model certain characteristics, we use TWS as a 672-dimensional feature space for  180 spatial observations (zipcodes). We drop part of the observations due to the sparse Twitter stream (leading to pikes of relative activity over 10\% of overall activity per zipcode) or missing values for target variables. 


In order to avoid over-fitting and optimize model hyper-parameters, the data and dependent variables are split into 3 parts, — training, testing and validation parts, represented by 60\%, 20\% and 20\% of all records correspondingly.

For each model, we apply a set of appropriate parameters, including kernel types, coefficients, number of trees, etc. We iterate through all possible combinations of those parameters, storing the model with best results over the validation set for each target variable. Best model is then applied to the test set for the final evaluation.

After assessing the out-of-sample performance for each model, we select the best model for each metric, and report its performance.

\end{document}